\documentclass[aps]{revtex4}

\usepackage{graphicx}
\usepackage{epsfig}
\usepackage{amsmath}
\usepackage{times}
\usepackage{bm}
\def\@biblabel#1{{#1.}} %    %  LV \def\@BIBLABEL#1{$^{#1}\m@th$} %
\def\be{\begin{equation}} \def\ee{\end{equation}}
\newcommand{\ket}[1]{\mbox{$|#1\rangle$}}

\begin{document}
\normalsize

%\begin{document}

\title{Channelization architecture for wide-band slow light in atomic vapors}

\author{Zachary Dutton, Mark Bashkansky, Michael Steiner and John Reintjes}
\affiliation{Naval Research Laboratory, Washington, DC 20375}

\begin{abstract}

We propose a ``channelization'' architecture to achieve wide-band
electromagnetically induced transparency (EIT) and ultra-slow light
propagation in atomic ${}^{87}$Rb vapors.  EIT and slow light are
achieved by shining a strong, resonant ``pump'' laser on the atomic
medium, which allows slow and unattenuated propagation of a weaker
``signal'' beam, but only when a two-photon resonance condition is
satisfied.  Our wideband architecture is accomplished by dispersing
a wideband signal spatially, transverse to the propagation
direction, prior to entering the atomic cell. When particular Zeeman
sub-levels are used in the EIT system, then one can introduce a
magnetic field with a linear gradient such that the two-photon
resonance condition is satisfied for each individual frequency
component.  Because slow light is a {\it group} velocity effect,
utilizing differential phase shifts across the spectrum of a light
pulse, one must then introduce a slight mismatch from perfect
resonance to induce a delay.  We present a model which accounts for
diffusion of the atoms in the varying magnetic field as well as
interaction with levels outside the ideal three-level system on
which EIT is based. We find the maximum delay-bandwidth product
decreases with bandwidth, and that delay-bandwidth product $\sim 1$
should be achievable with bandwidth $\sim 50$~MHz ($\sim 5$~ns
delay). This is a large improvement over the $\sim 1$~MHz bandwidths
in conventional slow light systems and could be of use in signal
processing applications.

\end{abstract}

\maketitle

%%%%%%%%%%%%%%%%%%%%%%%%%%%%%%%%%%%%%%%%%%%%%%%%%%%%%%%%%%%%%
\section{INTRODUCTION}
\label{sect:intro}

Electromagnetically induced transparency (EIT)
\cite{EIT,quantumOptics}, in which a ``pump'' field of laser light
can allow a weaker ``signal'' field to propagate through an
otherwise opaque atomic gas, has been inspiring a number of
applications based on the underlying coherent interaction of laser
light with atomic media. These include nonlinear optics at low light
levels \cite{NLoptics} and ultra-sensitive magnetic field
measurements \cite{magneticSensing}.

Of particular interest has been the recent observation of ultra-slow
light (USL) \cite{slowCold,slowThermal} in atomic gases, at group
velocities on the order of 10~m/s, due to a steep linear dispersion
in the index of refraction associated with the narrow EIT feature.
This could allow for controllable true-time delay devices for
classical light pulses, with applications in fiber-optic
telecommunications \cite{telecom} and radar signal processing
\cite{radar}.  Later extensions of the technique to stored light
\cite{stoppedCold,stoppedThermal} (for several milliseconds) has
also raised the possibility of quantum memory devices
\cite{quantumProcessing}. While the narrow frequency feature (below
the natural linewidth and Doppler width) of EIT is one of its
attractive features for precision applications
\cite{magneticSensing}, this has drawbacks in regards to delay and
storage applications. Optical communications and radar processing
typically desire $\sim 20$~GHz bandwidth. Similarly, single photon
sources and other tools of potential quantum information
technologies may emit photons over a broad band. In USL experiments
to date, the width of EIT transparency window is much narrower.

\begin{figure}
   \begin{center}
   \begin{tabular}{c}
   \includegraphics[height=3cm]{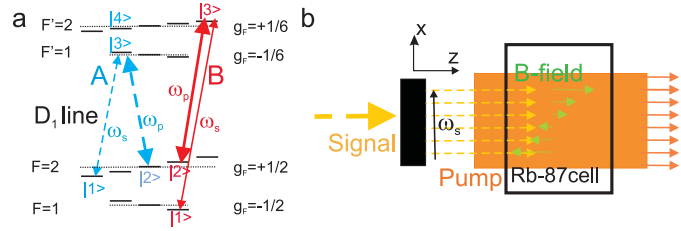}
   \end{tabular}
   \end{center}
   \caption[diagram]
   { \label{fig:diagram}
\textbf{Schematic of wideband atomic slow light system.}
\textbf{(a)} The hyperfine structure of the $D_1$ line in
${}^{87}$Rb.  The bare energy levels are shown as dotted lines.  The
excited state splitting is $(2 \pi) 817$~MHz, the ground state
splitting is $(2 \pi) 6.834$~GHz, and the $D_1$ optical transition
is at $\lambda=795$~nm \cite{Steck}.  A magnetic field along the
quantization axis ($z$) splits the individual Zeeman sub-levels
$\ket{i}$ (solid black lines) according to $\Delta_Z^{(i)} = \mu_B
g_F^{(i)} m_F^{(i)} B_z$ shown schematically in the figure for a
positive magnetic field. We consider two EIT $\Lambda$ schemes
(Scheme A as dashed arrows(blue online), Scheme B as solid arrows
(red online)). The thicker arrows indicate the pump field, with
frequency $\omega_p$ chosen to be resonant with the
$\ket{2}\leftrightarrow\ket{3}$ transition with no magnetic field.
The thinner arrows refer to the signal field, at some arbitrary
frequency $\omega_s$ within our desired bandwidth. The magnetic
field is chosen such so the system is in two-photon resonance. In
each case the atoms are initially purely in $\ket{1}$.  In Scheme A,
the fields also couple to $\ket{4}=\ket{F'=2,m_F=-1}$, causing a
decoherence mechanism which we investigate. \textbf{(b)} The
channelization architecture. A signal field is split, such that the
$x$ position of each frequency component is proportional to the
frequency, and then input into a cell illuminated by a monochromatic
pump field. A magnetic field with a linear gradient (dotted arrows,
green online) is then applied across the cell in such a way that two
photon resonance is nearly maintained everywhere.}
   \end{figure}

EIT and USL work best when the atom can be well described with a
$\Lambda$ energy level structure.  The signal field is near-resonant
with a stable state (which we label $\ket{1}$) and a radiatively
decaying excited state ($\ket{3}$).   The pump field is resonant
with another stable state $\ket{2}$ and the common excited level
$\ket{3}$.   We consider two energy level schemes  in ${}^{87}$Rb,
shown in Fig.\ref{fig:diagram}(a).  The schemes are labeled ``A''
(dashed, blue arrows) and ``B'' (solid, red arrows). The
transparency and slow, distortion-free propagation of the signal
pulse that we desire occur only when the frequency difference of the
two lasers $\omega_s-\omega_p$ matches the energy level difference
between levels $\ket{1}$ and $\ket{2}$ to within the narrow EIT
width. Frequency components of the signal outside this width are
strongly absorbed and distorted. This width is directly proportional
to the pump power and practical limits on the pump power ($\sim
10$~mW/cm$^2$) limit it to $\sim 1$~MHz.

The need for a wide-band, controllable true-time delay device
inspired us to here consider theoretically a ``channelization''
geometry utilizing Zeeman shifts in the atoms (see
Fig.\ref{fig:diagram}(b)). There are a variety of techniques
available to spatially separate various frequency components of
broadband light.  Assume a broadband signal pulse (represented as a
dashed, yellow arrow) propagating along the longitudinal ($z$)
dimension is split in the transverse ($x$) direction, with a
transverse displacement proportional to the detuning from some
chosen central frequency.  In our channelization geometry, this
dispersed signal then enters the ${}^{87}$Rb cell, which is
illuminated by a co-propagating, monochromatic pump field.  When
particular Zeeman sublevels are used, as in
Fig.\ref{fig:diagram}(a), the two photon resonance condition
required for slow light will be a strong function of magnetic field
along the quantization axis $z$ due Zeeman shifts of the levels.
Thus, applying a longitudinal magnetic field (dotted, green arrows)
with a linear gradient along $x$ will allow us to achieve the
conditions for EIT and USL for each frequency simultaneously. The
components can then be recombined after passing through the cell,
resulting in a true time delay device for a broadband signal.

USL is a {\it group} (rather than {\it phase}) velocity effect,
meaning it works by applying differential phase shifts to each
frequency component in the signal pulse. Maintaining perfect
two-photon resonance everywhere would result in no differential
phase shift and thus, no delay.  Therefore, one should choose the
magnetic field gradient such that there is a small, varying detuning
across the cell.  By choosing this mismatch to be small enough that
all components are within the EIT resonance one can obtain the
linear frequency dispersion necessary for slow light and still
maintain transparency.   There is a direct trade-off such that the
delay-bandwidth product cannot be increased with this method. But
for many applications, such as delays for signal processing, delays
in conventional slow light systems are much longer than necessary,
while the bandwidth is much too small.  Our method allows one to
circumvent this problem.

To be of practical interest, such a system would ideally provide
uniform transmission and delay over the entire bandwidth.  We will
see that Doppler broadening due to thermal motion of a room
temperature vapor is actually beneficial to our scheme in this
respect. Two-photon resonance is maintained for atoms of all
velocities because the two fields are co-propagating, while the
one-photon detuning strongly depends on the velocity of each atom.
Once one averages over the Doppler profile one finds that delay and
transmission insensitive to the one-photon detuning when it is well
within the Doppler width ($\sim$300~MHz in room temperature
rubidium) \cite{onePhoton}.  This means that the one-photon detuning
resulting from our Zeeman shifts (see Fig.~\ref{fig:diagram}(a))
does not effect our delays and transmissions.

There are several issues to consider to optimize the system we
propose here.  First, the transmission (and maximum delay) are
limited by decay of the coherence between the two ground states.
This decay time is often governed by the time it takes the atoms to
leave the interaction region with the pump (due to thermal motion)
\cite{slowThermal}. This decoherence mechanism can be mitigated by
the addition of a buffer gas (such as helium)
\cite{bufferCoherence,slowThermal,Arimondo} which significantly
reduces the diffusion of the ${}^{87}$Rb atoms via collisions.
Second, decoherence can also occur from transitions to levels
outside the $\Lambda$ system of interest (for example, the level
$\ket{4}$ in Scheme A in Fig.~\ref{fig:diagram}(a)).  The presence
of a high pressure of buffer gas can actually worsen this problem by
pressure broadening these unwanted levels.   This means there is
often an optimal intermediate pressure which balances these two
considerations.  A third consideration, unique to the channelization
architecture, is the presence of a high magnetic field gradient.
Because the two photon resonance condition is only satisfied over a
small range of magnetic fields, and therefore a small transverse
spatial region, the thermal diffusion of atoms will put them out of
two-photon resonance and potentially cause a severe absorption of
the signal. This will mean higher buffer gas pressures may be
desirable for our proposal than in conventional slow light.

In this paper, we present calculations of the transparency, delays
and bandwidths for both Schemes A and B in ${}^{87}$Rb atoms with a
helium buffer gas.  We use a model based on the linear response of
the signal in the medium, taking into account the diffusion of the
atoms, the linear gradient of the magnetic field, pressure
broadening, Doppler broadening, as well as interaction with
additional levels in the hyperfine structure. The calculation uses a
semi-classical model of the evolution of the ensemble average of the
atomic density matrix interacting with classical light fields, and a
classical treatment of the light field propagation based on the
slowly varying envelope (SVE) version of Maxwell's equations
\cite{quantumOptics}.

In the first part of the paper we consider the case of a spatially
uniform (but arbitrary) magnetic field and present a systematic
analysis and optimization of the transmission and delays (with
regards to buffer gas pressure and ${}^{87}$Rb density, etc.). In
the second part, we then introduce a model to account for a magnetic
field with a steep linear gradient.  We then discuss how to best
choose the slight mismatch in the two-photon resonance to maximize
the delay-bandwidth product.  We find that Scheme B, provides better
overall performance, with delays $\sim$5~ns over a bandwidth
$\sim$50~MHz.  We find that the delay-bandwidth product decreases
with bandwidth, reaching unity at around 50~MHz. However, it is more
difficult to prepare the initial state for this scheme (all atoms in
$\ket{F=1,m_F=+1}$) than it is for Scheme A (all atoms in
$\ket{F=2,m_F=-2}$, which is easily initialized via optical pumping
from the pump). While Scheme A provides worse overall performance,
it should still provide a suitable system to improve the bandwidth
over conventional systems.

\section{\label{sec:uniform} Uniform magnetic field case}

We begin by analyzing ``conventional'' slow light
\cite{slowCold,stoppedThermal}, with a homogenous magnetic field and
attempt to understand the delays, bandwidths and delay-bandwidth
products achievable in both schemes A and B. Our model accounts for
effects of an extra level $\ket{4}$ and the buffer gas and can also
explore the dependence on the magnetic field.

\subsection{\label{subsec:modelUniform} Model}

We model the signal and pump light fields classically and represent
them with Rabi frequencies $\Omega_{s(p)} \equiv - e
\mathbf{r}_{13(23)}\cdot \hat{\epsilon}_{s(p)} E_{s(p)}/\hbar$,
where $\mathbf{r}_{ij}$ are dipole matrix elements,
$\hat{\epsilon}_{s(p)}$ are unit polarization vectors, and
$E_{s(p)}$ are the slowly varying envelopes of the electric fields
$E_{s(p)}(e^{i(k_{s(p)} z - \omega_{s(p)} t)}+ c.c)/2$. Here the
wavenumbers $k_s,k_p \approx 2 \pi/\lambda$, with $\lambda = 795$~nm
the wavelength, are taken to be equal. The field polarizations
$\hat{\epsilon}_{s(p)}$ are chosen to match the transitions
($\sigma_+,\sigma_-$ in Scheme A and $\sigma_+,\sigma_+$ in Scheme
B). Meanwhile the ${}^{87}$Rb atoms are represented by a $4 \times
4$ density matrix $\hat{\rho}^{0}$ (representing each internal level
$\ket{i}$ under consideration). The diagonal elements represent the
fractional populations in each state while the off-diagonal terms
represent the coherences between levels (induced by the coherent
lasers). At the microscopic level, each atom has a density matrix
but we course grain average over spatial regions large compared to
the inter-atomic spacing, meaning the total density matrix can be
approximated by $N \hat{\rho}^{0}$, where $N$ is the atomic density
and $\hat{\rho}^{0}$ is normalized to unity $\sum_{i}
\rho^{0}_{ii}=1$.

Our model for the atomic density matrix evolution has two parts: (1)
a coherent, Hamiltonian part, which includes coupling with the light
fields, Zeeman shifts, and pressure shifts; and (2) an incoherent
part, which includes spontaneous decay, pressure broadening, and
diffusion out of the pump field interaction region (or the cell
walls if the pump illuminates the entire cell). The evolution
equations for the density matrix elements are given by:

\begin{eqnarray}
\label{eq:atomicEvol}\dot{\rho}_{ij}^0 & = &
 -(i/\hbar)[\hat{\rho}^0,\hat{\mathcal{H}}]_{ij}
+ \sum_{kl} \mathcal{L}^{ij}_{kl} \rho^0_{kl};  \\
\mathrm {where} \nonumber \\
\label{eq:H} \hat{\mathcal{H}}& = & \hbar \left(\begin{array}{cccc}
\omega_1 + \Delta_Z^{(1)}  &  0 & \frac{1}{2}\Omega_s^* e^{-i (k_s z
- \omega_s t)} & \frac{1}{2}\beta_{14}\Omega_s^* e^{-i (k_s z -
\omega_s t)} \\ 0 & \omega_2 +  \Delta_Z^{(2)}
 & \frac{1}{2}\Omega_p^* e^{-i (k_p z - \omega_p t)}
&  \frac{1}{2}\beta_{24}\Omega_p^* e^{-i (k_p z - \omega_p t)}\\
\frac{1}{2}\Omega_s e^{i (k_s z - \omega_s t)} & \frac{1}{2}\Omega_p
e^{i (k_p z - \omega_p t)}
&  \omega_3 +  \Delta_Z^{(3)} + S_p^{(\mathrm{e})} p & 0 \\
\frac{1}{2}\beta_{14}\Omega_s e^{i (k_s z - \omega_s t)}  &
\frac{1}{2}\beta_{24}\Omega_p e^{i (k_p z - \omega_p t)}   & 0 &
\omega_4 +  \Delta_Z^{(4)} + S_p^{(\mathrm{e})} p
\end{array} \right).
\end{eqnarray}

\noindent The parameters $\beta_{14(24)} =
\hat{\epsilon}_{s(p)}\cdot
\mathbf{r}_{14(24)}/\hat{\epsilon}_{s(p)}\cdot \mathbf{r}_{13(23)}$
characterize the difference in coupling (both sign and amplitude)
for the unwanted transition to $\ket{4}$. They are given by ratios
of Clebsch-Gordon coefficients and are
$\beta_{14}=1/\sqrt{3},\beta_{24}=-\sqrt{3}$ for Scheme A
\cite{Steck} and vanish for Scheme B, where there is no coupling to
additional levels. We have made the rotating wave approximation to
eliminate counter-rotating terms and coupling to levels detuned by
the ground-state hyperfine frequencies ($\sim 6.8$~GHz), but kept
terms detuned only by the excited-state hyperfine detuning ($\sim
800$~MHz).  The bare level frequencies $\omega_j$ on the diagonal
terms are shifted by linear Zeeman shifts $\Delta_Z^{(i)} = \mu_B
g_F^{(i)} m_F^{(i)} B_z$ where the Bohr magneton is $\mu_B = (2 \pi)
1.4~\mathrm{MHz/G}$, the Lande g-Factors $g_F^{(i)}$ are given in
Fig.~\ref{fig:diagram}(a), and $B_z$ is the magnetic field
\cite{Steck}.  They are also shifted by buffer gas pressure shifts
$S_p^{(i)} p$, where $p$ is the pressure. For ${}^{87}$Rb with a
helium buffer gas, these are taken to be $S_p^{(\mathrm{e})}=-(2
\pi) 0.9$~MHz/Torr for all excited manifold ($F'=1,F'=2$) states
\cite{expBuffer} (the ground state pressure shifts are much smaller
and negligible for our parameters).

The incoherent evolution is governed by a super-operator
$\tilde{\mathcal{L}}$. Writing out only the non-zero terms:

\begin{eqnarray}
\label{eq:L} \mathcal{L}^{11(22)}_{33}  &= &+ \Gamma_r f_{13(23)},
\hspace{2cm} \mathcal{L}^{11(22)}_{44}=  + \Gamma_r f_{14(24)},
\hspace{2cm}
\mathcal{L}^{33(44)}_{33(44)}   = -\Gamma_r  \nonumber \\
\mathcal{L}^{12}_{12} &=& -  \gamma_{\mathrm{diff}}, \hspace{2cm}
\mathcal{L}^{13(14)}_{13(14)}  = - (\Gamma_r/2 + B_p p),
\hspace{2cm} \mathcal{L}^{34}_{34}  =  - (\Gamma_r + 2 B_p p)
\end{eqnarray}

\noindent (plus the corresponding complex conjugate terms for the
off-diagonal elements). These terms are generally non-Hermitian as
they involve lossy terms and incoherent transitions due to
spontaneous emission.  The first line represents feeding the ground
states $\ket{1},\ket{2}$ via spontaneous emission from
$\ket{3},\ket{4}$.  The rate of emission from these levels is
$\Gamma_r= (2 \pi) 6$~MHz but this can branch into both levels
$\ket{1},\ket{2}$ as well as other levels outside of the system of
interest.  The branching ratios to various states are given by the
oscillator strengths $f_{ij}$ which are proportional to the square
of the Clebsch-Gordon coefficients.  For Scheme A, $f_{13}=1/2,
f_{23}=1/12$ and $f_{j4} = |\beta_{j4}|^2 f_{j3}$, while for Scheme
B, $f_{13}=1/2, f_{23}=1/6$.  The second line represents the
population loss of the excited states from spontaneous emission.  In
the third line, we have put in the dephasing of the ground state
coherence due to the diffusion of the gas out of the interaction
region (taken to be of width $w_\mathrm{int}$ and height
$h_\mathrm{int}$) \cite{Arimondo}:

\begin{equation}
\label{eq:gammaDiff} \gamma_{\mathrm{diff}}=\frac{2}{3}2.405^2
\frac{D_g}{w_\mathrm{int}h_\mathrm{int}}\frac{1}{1+6.8 \,
l_{\mathrm{mfp}}/\sqrt{w_\mathrm{int}h_\mathrm{int}}},
\end{equation}

\noindent where the diffusion constant for ${}^{87}$Rb in a helium
buffer gas of pressure $p$ and temperature $T$ is \\ $D_g =
(410~\mathrm{cm^2/s}) (\mathrm{Torr}/p)\sqrt{T/273~\mathrm{K}}$
\cite{Happer}. The mean free path is $l_{\mathrm{mfp}} = 3
D_g/v_{th}$ and the thermal velocity $v_{th}=\sqrt{3 k_B T/m}$. The
factor (2/3) in front is to account for the fact that length (along
$z$) is generally much longer than $w_\mathrm{int},h_\mathrm{int}$
and diffusion in this dimension does not effect the coherence. There
is additionally a depolarizing cross-section from ${}^{87}$Rb-He
collisions, which can dephase the ground states, but this is much
smaller effect than other decoherence mechanisms for our parameters
of interest and is neglected.  The last two lines of
Eq.~(\ref{eq:L}) contain dephasings of coherences from radiation, at
$\Gamma_r/2$, and pressure broadening, with $B_p=(2 \pi)$5~MHz/Torr
\cite{expBuffer,bufferAbs}.

When solving Eq.~(\ref{eq:atomicEvol}), we consider the usual weak
signal regime, and drop all terms higher than linear order in
$\Omega_s$. This is valid when $|\Omega_s| \ll |\Omega_p|$ and when
multiple scattering of spontaneously emitted photons can be ignored
(which is usually the case in EIT since spontaneous emission is
suppressed). Then we can take $\rho^0_{11} \rightarrow 1$ and
$\rho^0_{22},\rho^0_{33},\rho^0_{44},\rho^0_{23},\rho^0_{24},\rho^0_{34}
\rightarrow 0$. We are left with three non-trivial equations for the
evolution of $\rho^0_{21},\rho^0_{31},\rho^0_{41}$. We furthermore
make transformations to eliminate the time-dependent terms in
Eq.~(\ref{eq:H}): $\rho_{31,41} = \rho^0_{31,41} e^{-i (k_s z -
\omega_s t)},\rho_{21} = \rho^0_{21} e^{-i ((k_s-k_p) z
-(\omega_s-\omega_p) t)}$. For convenience we define a vector
$\mathbf{\rho} \equiv [\rho_{21},\rho_{31},\rho_{41}]^T$. The
evolution Eq.~(\ref{eq:atomicEvol}) can then be written as:

\begin{eqnarray}
\label{eq:atomicEvolLin} \dot{\mathbf{\rho}} & = &
\hat{\mathcal{M}} \mathbf{\rho} + \mathbf{S}; \nonumber \\
\hat{\mathcal{M}} & \equiv  & \left(\begin{array}{ccc} i
(\Delta_s'-\Delta_p') - \gamma_{\mathrm{diff}} & - \frac{i}{2}
\Omega_p^* & -\frac{i}{2} \beta_{24} \Omega_p^* \\
 -\frac{i}{2} \Omega_p & i (\Delta_s'+ \delta_D) - \gamma_e & 0\\
 - \frac{i}{2} \beta_{24} \Omega_p & 0 &
 i (\Delta_s' - \Delta_{43}' + \delta_D) - \gamma_e \nonumber \end{array} \right), \nonumber \\
\mathbf{S}  &\equiv & (0,-\frac{i}{2} \Omega_s,- \frac{i}{2}
\beta_{14} \Omega_s)^T
\end{eqnarray}

\noindent where we have defined the shifted detunings,
$\Delta_s'=\Delta_s - \Delta_Z^{(3)} - S_p^{(\mathrm{e})} p
+\Delta_Z^{(1)}, \, \Delta_p'=\Delta_p - \Delta_Z^{(3)} -
S_p^{(\mathrm{e})} p +\Delta_Z^{(2)}$, and $\Delta_{43}'
=\omega_4^{(0)}+ \Delta_Z^{(4)} -\omega_3^{(0)}-\Delta_Z^{(3)}$. The
bare detunings are $\Delta_{s(p)}=\omega_{s(p)}-(\omega_3
-\omega_{1(2)})$.  Throughout we choose pump to resonant with the
bare resonance $\Delta_p=0$, while the signal $\Delta_s$ varies. We
have also introduced a Doppler shift $\delta_D = (2 \pi)
v_z/\lambda$, where $v_z$ is the velocity of a particular atom along
the light propagation direction $z$. The total dephasing rates of
the optical transitions are $\gamma_e = \Gamma_r/2 + B_p p$.

When studying the light field propagation, it will be easiest to
work in Fourier space and so we transform from $\mathbf{\rho}(t)
\rightarrow \bar{\mathbf{\rho}}(\delta)$ and $\Omega_s(t)
\rightarrow \bar{\Omega}_s(\delta)$.
Equation~(\ref{eq:atomicEvolLin}) is linear in time-dependent
quantities and finding the solution of its Fourier transform is
equivalent to solving for its steady state in the time domain but
replacing $\Delta_s \rightarrow \bar{\Delta}_s = \Delta_s + \delta$
where $\delta$ represents the deviation of a particular Fourier
component of the signal field $\bar{\Omega}_s$ from the central
probe frequency $\omega_p$ (due to time dependence).  This solution
is:

\begin{equation}
\label{eq:rhoSS} \bar{\mathbf{\rho}} = \hat{\bar{\mathcal{M}}}^{-1}
\mathbf{\bar{S}},
\end{equation}

\noindent where $\hat{\bar{\mathcal{M}}}^{-1}$ is simply
$\hat{\mathcal{M}}^{-1}$ after making the replacement $\Delta_s
\rightarrow \bar{\Delta}_s$ and in $\mathbf{\bar{S}}$ we replace
$\Omega_s \rightarrow \bar{\Omega}_s$.

Finally, to obtain the response of the entire medium we integrate
over the thermal profile of velocities $v_z$ \cite{slowThermal}:

\begin{equation}
\label{eq:SuscDoppler} \bar{\rho}^{(D)}(\bar{\Delta}_s) = \int d
\delta_D \, \bar{\rho}(\bar{\Delta}_s,\delta_D) \,
\mathrm{exp}\bigg(-\frac{\delta_D^2}{\Delta_D^2}\bigg),
\end{equation}

\noindent where the Doppler width is $\Delta_D=\sqrt{2/3}(2
\pi)v_{th}/\lambda$ (the square root term is a geometrical factor).

Turning now to the light propagation, in the linear signal regime
the pump field Rabi frequency $\Omega_p$ is constant in space and
time. The signal field $\Omega_s$ propagates according to the SVE
Maxwell equation, with the polarization written in terms of the
atomic density matrix.  Once we Fourier transform the Maxwell
equation, it can be written:

\begin{eqnarray}
\label{eq:MaxwellSusc}\frac{\partial}{\partial z} \bar{\Omega}_s & =
& \frac{i}{2}N f_{13} \sigma \chi^{(D)}(\bar{\Delta}_s)
\bar{\Omega}_s + i \frac{\delta}{c} \bar{\Omega}_s
\nonumber \\
\mathrm{where} \hspace{1cm} \chi^{(D)}(\bar{\Delta}_s) & = &
-\frac{\Gamma_r}{\bar{\Omega}_s}(\bar{\rho}_{31}^{(D)}+ \beta_{14}
\bar{\rho}_{41}^{(D)})
\end{eqnarray}

\noindent and $\sigma = 3 \lambda^2/(2 \pi)$ is the resonant cross
section for unity oscillator strength.  The last term accounts for
the propagation at $c$ in free space.  If the fastest time scale of
interest in the problem is much slower than total length of
propagation ($\sim$~cm) divided by $c$ then this term can be
neglected.  Note that since $\bar{\rho}_{31},\bar{\rho}_{41} \propto
\bar{\Omega}_s$, the susceptibility is independent of
$\bar{\Omega}_s$. Equation~(\ref{eq:MaxwellSusc}) can be trivially
solved to give how a particular frequency component, with an input
amplitude $\bar{\Omega}_s^{(\mathrm{in})}(\bar{\Delta}_s)$, will
propagate through a cell of length $l_\mathrm{cell}$ in the medium
$\bar{\Omega}_s^{(\mathrm{out})}(\bar{\Delta}_s)  = e^{\frac{i}{2}D
\chi^{(D)}(\bar{\Delta}_s)}
 \bar{\Omega}_s^{(\mathrm{in})}(\bar{\Delta}_s)$,
where the ``optical density'' is defined as $D = f_{13} N \sigma
l_\mathrm{cell}$ (we have ignored the speed of light propagation
term for simplicity). From this one clearly sees how the imaginary
part of $\chi^{(D)}$ is proportional to the absorption cross-section
while the real part is gives phase shifts and thus determines the
index of refraction.

In a non-Doppler broadened, non-pressure broadened medium in the
absence of a pump field ($\Omega_p \rightarrow 0$), and with no
level $\ket{4}$ ($\beta_{14}=0)$, we recover the usual Lorentzian
susceptibility profile.  The absorption cross-section is peaked at
the atomic resonance $\Delta_s'=0$ and has a width $\Gamma_r$ while
the real part exhibits anomalous dispersion at the resonance.  The
dimensionless susceptibility $\chi^{(D)}$ in
Eq.~(\ref{eq:MaxwellSusc}) has been defined such that the peak
absorption is $\chi^{(D)}(\Delta_s'=0) = i$, leading to exponential
attenuation of the signal intensity $I_s \propto e^{-D}$. Pressure
broadening and Doppler broadening act to reduce this resonant cross
section while widening the feature.

\subsection{\label{subsec:resultsUniform} Results}

The presence of a sufficiently strong pump field $\Omega_s$ will
then introduce a sharp EIT feature at frequencies very near the
two-photon resonance, in the middle of this broad absorption
resonance.  At the EIT resonance, we have reduced absorption
(transparency) and a linear slope in the index of refraction with
normal (positive) dispersion (leading to the slow group velocity).
We now proceed with some  calculations of these EIT features. The
primary motivation of the present section is to learn the role
played by the buffer gas pressure and to quantitatively learn the
dependence on magnetic field.  These results will later guide the
optimization of the channelization architecture with the linear
magnetic field gradient.

Throughout this section consider a case with a 1~mW pump laser with
an area $w_\mathrm{int}h_\mathrm{int}= (2~\mathrm{mm})^2$. Due to
differing oscillator strengths $f_{23}$ this results in Rabi
frequencies of $\Omega_p=(2 \pi)8.45$~MHz for Scheme A and
$\Omega_p=(2 \pi)12.3$~MHz for Scheme B.  At a temperature of
$T=333$~K (60 degrees Celcius), there is a ${}^{87}$Rb density of
$N=2.5 \times 10^{11}~\mathrm{cm}^{-3}$ \cite{Steck} and a Doppler
width $\Delta_D=(2 \pi) 320$~MHz.

\begin{figure}
   \begin{center}
   \begin{tabular}{c}
   \includegraphics[height=2.5cm]{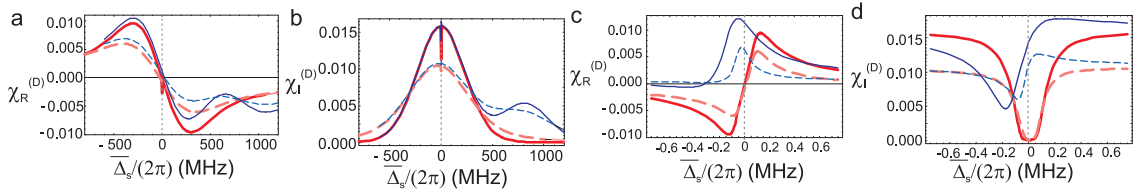}
     \end{tabular}
   \end{center}
\caption{\label{fig:chiScheme} \textbf{EIT transparency window and
steep index of refraction.} \textbf{(a-b)} Frequency dependence of
the real (a) and imaginary (b) parts of the susceptibility
$\chi^{(D)}$ in Schemes A (thin curves, red online) and B (thick
curves, blue online) with $B_z=0$. Buffer gas pressures are p=3 Torr
(solid curves) and 30 Torr (dotted curves).  The scale is too large
to see the EIT resonance features, which is located at the vertical
dotted lines. \textbf{(c-d)} The same curves shown on a smaller
scale, at the EIT resonance.  The dotted lines indicate exact
two-photon resonance $\Delta_p'=\Delta_c'$.}
\end{figure}

Considering first a homogenous $B_z=0$ magnetic field, we plot in
Fig.\ref{fig:chiScheme} the real and imaginary parts of the Doppler
averaged susceptibility
$\chi_R^{(D)}(\bar{\Delta}_s),\chi_I^{(D)}(\bar{\Delta}_s)$,
according to Eq.~(\ref{eq:MaxwellSusc}), both on a large (a,b) and
small (c,d) scale for both Schemes A (thinner, blue curves) and
Scheme B (thicker, red curves). From Fig.\ref{fig:chiScheme}(a,b) we
see that, away from the narrow EIT feature, the susceptibility
retains the usual Lorentzian susceptibility feature one would expect
in a two-level medium, with the width $\sim \Delta_D$ and height
$\sim \Gamma_r/2\Delta_D$. The biggest apparent difference between
the two schemes is that in Scheme A there is an extra resonance, due
to level $\ket{4}$, near the hyperfine splitting ($\sim 817$~MHz).
Note that we are in a regime where the Doppler broadening is
comparable to this splitting, causing the two resonances to slightly
overlap. The solid curves are for the relatively small buffer gas
pressure $p=3$~Torr, while the dotted curves are for $p=30$~Torr. At
this higher pressure, the pressure broadening becomes important as
$\gamma_e = (2 \pi)~253$~MHz becomes comparable to $\Delta_D$.

Examining the smaller scale plots Fig.~\ref{fig:chiScheme}(c-d), we
see the pump field introduces a sharp feature at the two-photon
resonance ($\bar{\Delta}_s=0$). In the imaginary part we see a
narrow notch in the absorption cross-section (the transparency
window) while in the real part we see steep linear dispersion, the
origin of the slow group velocity.  To describe the degree of
transparency, we define the parameter $R_{\mathrm{EIT}}$ as the
ratio of the minimum of $\chi_I^{(D)}(\bar{\Delta}_s)$ in the
resonance to its value there with no pump field present.   We note a
very large difference in this parameter for the two schemes, with
only $R_{\mathrm{EIT}} \approx 1/3$ for Scheme A. As we will discuss
below, coupling to level $\ket{4}$ is the reason for the lack of
good transparency in this case.  We also note that the transparency
window is slightly shifted from exact two photon resonance (by
approximately $-0.1~$MHz), an effect of stark shifts from $\ket{4}$.
In Scheme B, the transparency is quite good $R_{\mathrm{EIT}} \sim
0.001$.

To translate these $\chi^{(D)}(\bar{\Delta}_s)$ curves into
performance for delay devices, we examine the shape of these curves
at the EIT resonance.  To a good approximation, the real part here
can be written as some central value plus a linear part, while the
imaginary part is some minimum value plus a parabolic shape:

\begin{equation}
\label{eq:suscRes} \chi^{(D)}(\bar{\Delta}_s) \approx \phi_0 + S
(\bar{\Delta}_s-\Delta_0) + i A + i
\frac{(\bar{\Delta}_s-\Delta_0)^2}{W^2}
\end{equation}

\noindent where $\Delta_0$ is defined to be the minimum point of
$\chi_I^{(D)}$ and the parameters $\phi_0,\,A \, \mathrm{absorption}
\, S \, \mathrm{(slope)},\, W \, \mathrm{(width)}$ are obtained by
numerically evaluating $\chi^{(D)}(\bar{\Delta}_s)$ and its
derivatives at $\Delta_0$.  In these terms our transparency
parameter is
$R_\mathrm{EIT}=A/\mathrm{Im}\{\chi^{(D)}(\Delta_0)|_{\Omega_p
\rightarrow 0} \}$.

Now suppose we input a signal pulse a central frequency
$\Delta_s=\Delta_0$ and a $1/e$ intensity half-width $\tau_s$
(giving it frequency components in the range
$\bar{\Delta}_s=\Delta_0 \pm \tau_s^{-1}$). To calculate what the
pulse looks like after propagation through the medium, one takes the
Fourier transform of the input pulse, calculates the propagation of
each Fourier component according to Eq.~(\ref{eq:MaxwellSusc}), then
inverts the Fourier transformation back to the time domain.  If
$\tau_s$ sufficiently long that Eq.~(\ref{eq:suscRes}) is valid for
all frequency components of the pulse (which is usually true when
$\tau_s^{-1} \ll W$), then one can analytically perform the inverse
transformation. The result is a pulse which is delayed in time by
$\tau_D = D S/2$ and attenuated by a factor $e^{-A D}$. This Fourier
analysis of the propagation reveals clearly how the delay comes
about by a differential phase shift of the different frequency
components $S \Delta_s' D/2$.  One immediately sees there is a
trade-off between transparency and delay.  In addition, frequency
components slightly off the resonance will be preferentially
absorbed. This leads to a time-broadening of the pulse by $\tau
\rightarrow \sqrt{\tau_s^2+\beta^{-2}}$, where $\beta=W/\sqrt{D}$,
and a reduction in the peak intensity by
$\tau^2/(\tau_s^2+\beta^{-2})$. In this sense, $\beta$ can be
interpreted as the bandwidth of the system, as $\delta \sim
\tau_s^{-1} \ll \beta$ is required to prevent attenuation and
distortion of the pulse.

An obvious figure of merit is the delay-bandwidth product, important
for many applications such as quantum memory devices or optical
buffers.  This parameter indicates the degree of ``pulse
separation'' one can achieve ({\it i.e.} the number of pulse widths
one can delay). Incidentally, this parameter being unity corresponds
to the point at which the differential phase shifts across the
frequency spectrum of the pulse is $\sim (2 \pi)$. However, the
absolute value of the bandwidth and maximum delay are also of
importance, depending on the application.   All these parameters
depend strongly on the optical density $D$, which can generally be
adjusted since the ${}^{87}$Rb density $N$ is a strong function of
temperature \cite{Steck}.  At $N=2.5 \times
10^{11}~\mathrm{cm}^{-3}$ for a $l_\mathrm{cell}=1$~cm cell,
$D=370$.  For a given desired transmission, the parameter $A$ tells
us the optical density through which a pulse can successfully
propagate.  For a transmission $1/e$ we need
$D<D_\mathrm{max}=A^{-1}$.  This, in turn, determines the maximum
achievable delay $\tau_D^{\mathrm{(max)}}=S/2A$. The delay-bandwidth
product is $\tau_D \beta=S W \sqrt{D}/2$. This product increases
with $D$ and the best possible delay-bandwidth (using
$D_{\mathrm{max}}$) is $S W/2 \sqrt{A}$.

\begin{figure}
   \begin{center}
   \begin{tabular}{c}
   \includegraphics[height=2.5cm]{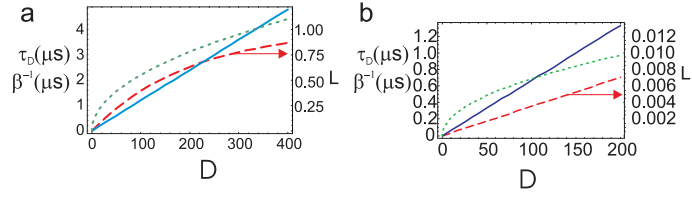}
     \end{tabular}
   \end{center}
\caption{\label{fig:propagation}\textbf{Pulse propagation
characteristics} \textbf{(a)} For the case $p=2$~Torr, $B_z=0$ in
Scheme A we calculate $A=0.005, \, S=24~\mathrm{ns}$, and $W=(2 \pi)
0.72$~MHz.  Here we plot the resulting delay $\tau_D=S D/2$ (solid
curve, blue online), loss $L=(1-e^{-A D}))$ (dashed, red) and
inverse bandwidth $\beta^{-1}=\sqrt{D}/W$ (dotted, green) versus
optical density $D$. \textbf{(b)} Same for Scheme B, with
$p=15$~Torr, for which $A=3.5 \times 10^{-5}$, $S=13.3$~ns, and
$W=(2 \pi) 2.31$~MHz.}
\end{figure}

For Scheme A we find the performance is better for lower pressures.
At $p=2$~Torr and other parameters as in Fig.~\ref{fig:chiScheme},
we calculate $A=0.005$, $S=24$~ns and $W=(2 \pi) 0.72$~MHz, giving a
maximum delay-bandwidth product of 0.76. In
Fig.~\ref{fig:propagation}(a), we plot resulting pulse propagation
parameters versus $D$. We see that for $D \approx 50$ one obtains a
$\tau_D=0.5~\mu$s delay with only 20\% loss.  But the minimum pulse
width here is $\beta^{-1}=1.5~\mu\mathrm{s}>\tau_D^{\mathrm{max}}$
One does not get into the ``pulse separation'' regime until at about
$D=350$, where $\tau_D =\beta^{-1}=4.1~\mu$s where the attenuation
is 82\% and the bandwidth is $\beta=(2 \pi)38$~kHz.

By contrast, we can do much better in Scheme B.  We find this system
is better at higher pressures and for $p=15$~Torr case we calculate
$A=3.5 \times 10^{-5}$, $S=13.3$~ns, and $W=(2 \pi) 2.31$~MHz,
giving $\beta \tau_D^{\mathrm{(max)}}=31$.  The pulse propagation
parameters are plotted in Fig.~\ref{fig:propagation}(b) and we see
the pulse separation regime begins at $D \approx 100$, at which
point $\tau_D \approx 0.7~\mu$s and the loss is only $L \approx
0.003$. The bandwidth here is $\beta=(2 \pi)230$~kHz. Note that
despite the better performance, one would still have to go to very
small $D$ to get the bandwidth significantly more than 1~MHz, and
here the delay-bandwidth product is quite small.

We now attempt to get some intuitive understanding of what
determines $A,\, S$, and $W$ in the two schemes and study the
dependence on buffer gas pressure and magnetic field. Scheme B,
because of the absence of level $\ket{4}$ is significantly easier to
understand. One can analytically obtain the solution
Eq.~(\ref{eq:rhoSS}) and plug it directly into the susceptibility in
Eq.~(\ref{eq:MaxwellSusc})). The conditions necessary for EIT are
that the ground state decoherence rate is small
$\gamma_{\mathrm{diff}} \ll \gamma_e$ and the pump field intensity
is sufficiently strong $|\Omega_p|^2 \gg \gamma_{\mathrm{diff}}
\gamma_e$.  Assuming these inequalities and Taylor expanding in
$\bar{\Delta}_s'$ about one- and two-photon resonance
($\bar{\Delta}_s'=\Delta_p'=0)$ one obtains a susceptibility in the
form of Eq.~(\ref{eq:suscRes}) with the parameters $\phi_0=0, \,
A=2\gamma_\mathrm{diff} \Gamma_r/|\Omega_p|^2, \, S=2
\Gamma_r/|\Omega_p|^2, \, W=|\Omega_p|^2/\sqrt{8 \gamma_e
\Gamma_r}$. Thus we find that the absorption scales as the inverse
of the pump intensity, the bandwidth scales directly with the
intensity and the delay scales inversely with the intensity. There
is a trade-off between delay and attenuation in choosing the pump
intensity (just like the optical density). However, unlike with
optical density, one cannot improve the delay-bandwidth product by
changing the pump intensity. One can increase the bandwidth (and
reduce the delay) but extremely high bandwidths require unreasonably
high pump powers. We find that these analytic expressions are still
valid after Doppler averaging. For the case we have just considered
(Fig.~\ref{fig:propagation}(b)) these estimates yield $A=3.5 \times
10^{-5}, S=13.3~\mathrm{ns}, W=(2 \pi)2.34$~MHz, in excellent
agreement with the numerical results.

At the point of transparency we find the ground state coherence is
$\rho_{21} \approx -\Omega_s/\Omega_p$, which is known as a ``dark
state'' \cite{EIT}. The strong pump field acts to drive the system
into this state.  In this case the two terms driving absorption, on
the $\ket{1} \leftrightarrow \ket{3}$ and $\ket{2} \leftrightarrow
\ket{3}$ transitions, (see the first two terms on the third line of
the Hamiltonian, Eq.~(\ref{eq:H})), are equal and opposite, leading
to a quantum interference which suppresses the absorption process.
However, a non-zero detuning $\Delta_p'$ or decoherence
$\gamma_{\mathrm{diff}}$ causes $\rho_{21}$ to slowly evolve out of
the dark state.  Our analytic expression $A=2 \gamma_\mathrm{diff}
\Gamma_r/|\Omega_p|^2$ reflects the steady-state which occurs due to
the balance of the preparation of the dark state by the pump field
and the loss from it due to diffusion.  To minimize our absorption,
we clearly want to minimize $\gamma_{\mathrm{diff}}$, which can be
done by increasing the buffer gas pressure (which decreases $D_g$)
or increasing the effective area of interaction $h_{\mathrm{int}}
\times w_{\mathrm{int}}$ (see Eq.~(\ref{eq:gammaDiff})).  In
practice, there is a trade-off between the interaction area and
$|\Omega_p|^2$ as one can increase the pump intensity by focusing
the it more tightly. Thus, for a given pump power and buffer gas
pressure, there is only a marginal dependence of $A$ on the focusing
area. Numerically we fine that there is a slight benefit in tighter
focusing. Regardless one can increase the pressure to improve $A$ to
the desired level.  The slope $S$ is almost completely unaffected by
$p$, while $W$ decreases with pressure due to the factor
$\sqrt{\gamma_e}$ in the denominator.

For Scheme A, the situation is significantly altered by the presence
of $\ket{4}$.  The problem is that the dark state  with respect to
absorptions into $\ket{3}$ (the state for which the two absorption
channels have equal and opposite amplitudes) is
$\rho_{21}=-\Omega_s/\Omega_p$ while the dark state with respect to
$\ket{4}$ is $\rho_{21}=-\beta_{14}\Omega_s/\beta_{24}\Omega_p$.
Thus, unless oscillator strength ratios are equal
$\beta_{14}=\beta_{24}$ there will be some absorption present for
any value of $\rho_{21}$.  In Scheme A, $\beta_{14}=1/\sqrt{3}$ and
$\beta_{24}=-1\sqrt{3}$.

This problem was studied in detail for the cold atom (non-Doppler
broadened) case in \cite{thesis}.  There it was found that, for
atoms nearly resonant with $\ket{3}$, this effect leads to minimum
absorption coefficient
$A_{\mathrm{off-res}}=(\beta_{14}-\beta_{24})^2 \gamma_e
\Gamma_r/4(\Delta_{43}'^2+\gamma_e^2)$ and an AC Stark shift of the
EIT resonance by $\Delta_{\mathrm{AC}} =
\beta_{24}(\beta_{24}-\beta_{14})|\Omega_p|^2\gamma_e/4
(\Delta_{43}'^2+\gamma_e^2)$.  In the Doppler broadened case this
problem is further complicated by the fact that, when the Doppler
width $\Delta_D$ is comparable to the hyperfine splitting
$\Delta_{43}'$, there is a significant fraction of atoms which
interact with both $\ket{3}$ and $\ket{4}$ with similar strength.
The EIT interference is almost completely destroyed for these atoms.
Numerically, we find that for Scheme A, this increases
$A_{\mathrm{off-res}}$ by nearly an order of magnitude from the
analytic estimate above.  Increasing pump power increases the
coupling to $\ket{4}$ in such a way that it exactly offsets any gain
in the strength of the EIT resonance, and so we find
$R_{\mathrm{EIT}}$ quickly saturates to a value around 0.3 (as in
Fig.~\ref{fig:chiScheme}(c-d)), a constant determined by the
relative values of $\Delta_D$ and $\Delta_{43}'$.  The interesting
thing to note about $A_{\mathrm{off-res}}$ is that it increases with
buffer gas pressure (via $\gamma_e$) because pressure broadening
increases the relative role played by coupling to $\ket{4}$.
Therefore, the diffusion problem favors higher $p$, while the
interaction with $\ket{4}$ generally favors lower pressure and the
lowest overall absorption is achieved by balancing these two
considerations. We find that the optimal pressure is rather low
(between 0.5 - 2 Torr, depending on interaction area), but that the
dependence is rather weak and so $R_{\mathrm{EIT}} \approx 0.3$ is a
good estimate over a broad range of pressures and interaction areas.
The AC Stark shift is visible in the plots
Fig.~\ref{fig:chiScheme}(c-d) and agrees well with the above
expression.

\begin{figure}
   \begin{center}
   \begin{tabular}{c}
   \includegraphics[height=2.5cm]{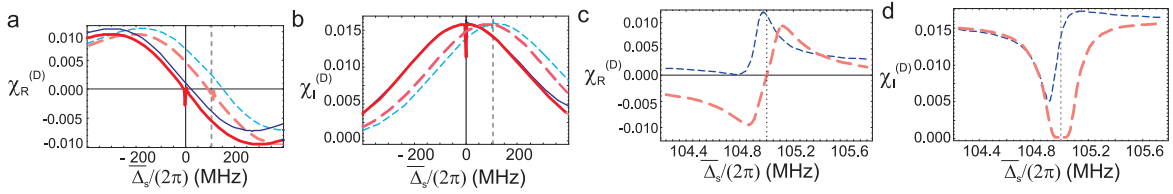}
     \end{tabular}
   \end{center}
\caption{\label{fig:chiB}\textbf{EIT in a homogenous magnetic
field.}  Plot of $\chi^{(D)}$ in the two schemes, again for
$p=3$~Torr. The solid curves are for $B_z=0$ (identical to
Fig.~\ref{fig:chiScheme}), while the dotted curves are with a
magnetic field $B_z=75$~G, which shifts the two-photon resonance
($\Delta_s'=\Delta_p'$) to $s_\mathrm{res} B_z=(2 \pi) 105$~MHz.}
\end{figure}

Finally, in preparation for our channelization calculations, we must
consider the degree to which our results depend on the value of the
homogenous magnetic field $B_z$.  We still choose our pump field to
be resonant with the bare $\ket{2} \leftrightarrow \ket{3}$
transition $\Delta_p=\omega_p-(\omega_3-\omega_2)=0$. This system
will then be in two-photon resonance for a probe photons with bare
detuning
$\Delta_s=\Delta_Z^{(2)}-\Delta_Z^{(1)}-\Delta_{\mathrm{AC}}
=s_\mathrm{res} B_z-\Delta_{\mathrm{AC}}$, where $s_\mathrm{res}
\equiv \mu_B(g_F^{(2)} m_F^{(2)}-g_F^{(1)} m_F^{(1)})$. In both
Schemes A and B $s_\mathrm{res}=(2 \pi)1.4$~MHz/G. In
Fig.~\ref{fig:chiB} we plot $\chi^{(D)}$ for both $B_z=0$ (solid
curves) (as in Fig.~\ref{fig:chiScheme}) and with a large magnetic
field $B_z=75$~G (dotted curves). In Fig.~\ref{fig:chiB}(a-b) we see
that the magnetic field shifts the overall one-photon detuning so
the Doppler broadened Lorentzian resonances are shifted. It also
shifts the two-photon resonance (vertical lines) by a slightly
different amount (so the EIT resonance with $B_z=75$~G is not
exactly at peak of the Lorentzian resonance). However, this
difference is still well within the Doppler width. Examination of
the $g_F$ and $m_F$ shows that the signal detuning from $\ket{3}$,
$\Delta_s'$, (see Eq.~(\ref{eq:atomicEvolLin}) and below) is only
$1/6$ of that of the two-photon resonance shift $s_\mathrm{res} B_z$
(this is true in both Schemes A and B, though the relative sign of
the shift for the two schemes is opposite). Thus, even a shift of
the two-photon resonance $s_\mathrm{res} B_z \sim \Delta_D$ will
result in substantially smaller one-photon detuning. This is
beneficial as the widths and strengths of the EIT resonances are
effected when the one-photon detuning becomes comparable to the
Doppler width \cite{onePhoton}. Comparison of plots of the EIT
resonance in Fig.~\ref{fig:chiB}(c-d) with the $B_z=0$ case in
Fig.~\ref{fig:chiScheme}(c-d) reveals they looks almost identical.

\begin{figure}
   \begin{center}
   \begin{tabular}{c}
   \includegraphics[height=2.5cm]{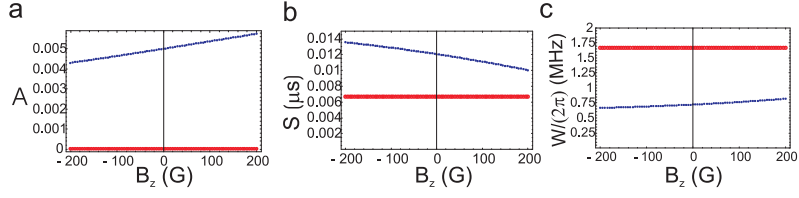}
     \end{tabular}
   \end{center}
\caption{\label{fig:versusB}\textbf{Dependence of EIT resonance on
magnetic field.} The parameters $A$ (a) $S$ (b) and $W$ (c) , versus
magnetic field $B_z$.  For Scheme A (thinner, blue), the pressure is
$p=1.2$~Torr, and for Scheme B (thicker, red) $p=30$~Torr. }
\end{figure}

To check numerically that the EIT resonance is indeed insensitive to
$B_z$ over a wide range, we calculated the parameters $A,S$ and $W$
over the range $B_z=$-200~G to +200~G.  The results, presented in
Fig.~\ref{fig:versusB} bear out our expectation.  In Scheme B there
is no visible dependence on $B_z$ on the scale plotted. There is a
weak dependence in the parameters in Scheme A, again primarily due
to the effect of level $\ket{4}$. The total range of resonance
frequencies for this range of magnetic fields is $(2\pi)560$~MHz.

\section{\label{sec:varying} EIT and slow light with a magnetic field gradient}

We are now prepared to add a final piece of our model account for
the large gradients in the magnetic field and the transverse spatial
dispersion of the signal in the proposed channelization geometry. We
will then use this to characterize the performance of the delay
device in each scheme, and explore quantitatively the maximum delays
versus bandwidth and the optimal buffer gas pressures.

\subsection{\label{subsec:modelGradient} Model}

Let us first consider how the susceptibility is affected by the
diffusion of the gas in the presence of a magnetic field gradient.
When a gas diffuses with some diffusion constant, it's density
matrix evolves as $\dot{\rho} = \cdots + D_g \nabla^2 \rho$
\cite{Steck} so we add this term to our existing evolution equation.
Since the magnetic field gradient only exists in the $x$ dimension,
the diffusion in $y$ and $z$ has no effect. We then write
Eq.~(\ref{eq:atomicEvolLin}):

\begin{equation}
\label{eq:atomicEvolDiff} \dot{\mathbf{\rho}}  = \hat{\mathcal{M}}
\mathbf{\rho} + \mathbf{S} +
\frac{D_g}{3}\frac{d^2}{dx^2}\mathbf{\rho}.
\end{equation}

\noindent With a magnetic field gradient our shifted detunings
$\Delta_s',\Delta_p'$ becomes spatially dependent and we take
$\frac{d^2}{dx^2}\mathbf{\rho} \rightarrow (\frac{dB_z}{dx})^2
\frac{\partial^2}{\partial B_z^2}\mathbf{\rho}$. Suppose we apply a
linear gradient so $B_z(x)=S_B x$ over the interaction region of
width $w_\mathrm{int}$.  Then the two-photon resonance would vary by
$\beta_\mathrm{eff} \equiv S_B w_\mathrm{int} s_\mathrm{res}$ over
the width of the interaction region, determining the effective
bandwidth of our system \footnote{We are assuming purely linear
Zeeman shifts throughout this paper, which is reasonable for shifts
much smaller than the ground state hyperfine splitting of $(2
\pi)6.8$~GHz. One could also account for the small quadratic shift
by introducing a small quadratic dependence in $B_z$ to
compensate.}.

Our susceptibility in the presence of diffusion will then be the
steady state solution of Eq.~(\ref{eq:atomicEvolDiff}).   It is
difficult to find this analytically in general, but this can be
achieved with a perturbation approach under the assumption that the
new diffusion term is small.  In the perturbative approach, we take
a zeroeth order solution $\mathbf{\bar{\rho}}^{(0)}$ to be solution
without the diffusion term
$\mathbf{\bar{\rho}}^{(0)}=\hat{\mathcal{\bar{M}}}^{-1}\mathbf{\bar{S}}$
(see Eq.~(\ref{eq:rhoSS})), and then plug this back into the full
equation to obtain a correction from the diffusion term:

\begin{equation}
\label{eq:rhoDiff} \mathbf{\bar{\rho}^{(1)}} =
S_B^2\frac{D_g}{3}\hat{\mathcal{M}}^{-1}\frac{\partial^2}{\partial
B_z^2}\mathbf{\bar{\rho}}^{(0)}
\end{equation}

\noindent Because the two-photon detuning $\Delta_s'-\Delta_p'$ is
directly proportional to $B_z$, the second derivative with respect
to the $B_z$ basically corresponds to the curvature with frequency
of $\chi^{(D)}$ at the EIT resonance, which Eq.~(\ref{eq:suscRes})
predicts to be $2/W^2$.  In a sense, the diffusion term causes an
averaging over some frequency width, which will partly wash out the
EIT resonance, increasing our minimum absorption $A$.  It also has a
tendency to widen the feature, increasing $W$ and decreasing our
$S$.

After obtaining our corrected density matrix
$\mathbf{\bar{\rho}}=\mathbf{\bar{\rho}^{(0)}}+\mathbf{\bar{\rho}^{(1)}}$
we once again Doppler average according to
Eq.~(\ref{eq:SuscDoppler}) and then calculate the susceptibility.
Unfortunately, the perturbative procedure is not valid near the
wings of the Doppler profile, where the EIT feature becomes very
narrow, even in cases where the perturbation is small near the
center of the Doppler profile. In the limit that the diffusion
correction becomes large $\mathbf{\bar{\rho}}$ should smoothly
return to it's value without the EIT feature
$\mathbf{\bar{\rho}}|_{\Omega_p\rightarrow 0}$, but in
Eq.~(\ref{eq:rhoDiff}) the correction $\bar{\rho}^{(1)}$ can grow
without bound.  For this reason, we must use a slightly more
complicated procedure, which in the limit of a small diffusion term
reproduces Eq.~(\ref{eq:rhoDiff}) and in the opposite limit reverts
to $\mathbf{\bar{\rho}}|_{\Omega_p\rightarrow 0}$.  This is
accomplished by taking an average over a small range of magnetic
fields:

\begin{eqnarray}
\label{eq:rhoDiffInt} \mathbf{\bar{\rho}}(\bar{\Delta}_s,\delta_D,B)
& = & \int d\delta_B
\frac{1}{\sqrt{\pi}\Delta_B}\mathbf{\bar{\rho}^{(0)}}(\bar{\Delta}_s,\delta_D,B_z+\delta_B)
\mathrm{Exp}\bigg(-\frac{\delta_B^2}{\Delta_B^2}\bigg) \nonumber \\
\mathrm{where} \hspace{1cm} \Delta_B^2 & = & \frac{4 D_g
S_B^2}{3}\bigg|\frac{\sum_j \lambda_j a_j \mathbf{v}_j}{\sum_j
\lambda_j \mathbf{v}_j}\bigg|,
\end{eqnarray}

\noindent and the $\{\lambda_j\}$ and $\{\mathbf{v}_j\}$ are,
respectively, the eigenvalues and eigenvectors of
$\hat{\bar{\mathcal{M}}}^{-1}$ and the ${a_j}$ are the coefficients
$a_j=\mathbf{v}_j\cdot(\partial^2 \mathbf{\bar{\rho}}^{(0)}/\partial
B_z^2)$. After calculating Eq.~(\ref{eq:rhoDiffInt}) we can then
Doppler average with Eq.~(\ref{eq:SuscDoppler}).

\begin{figure}
   \begin{center}
   \begin{tabular}{c}
   \includegraphics[height=2.1cm]{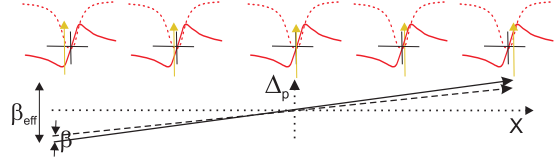}
     \end{tabular}
   \end{center}
\caption{\label{fig:widebandSchematic}\textbf{Schematic of detunings
in wideband slow light scheme.} The signal pulse frequencies are
dispersed along $x$ with a slope $S_\mathrm{disp}$, indicated by the
black solid arrow, which is slightly larger than the slope of the
two-photon resonance $S_B s _\mathrm{res}$, indicated with the
dashed arrow.  The values chosen for the mismatch in slopes is such
that the detuning local resonance $\Delta_p'$ varies linearly with
$x$ and is everywhere within the local bandwidth $\beta$. The top
row diagrams this mismatch with respect to the local resonance at
several locations, with the solid and dotted curves showing,
respectively, the general shape of the $chi^{(D)}_R,\chi^{(D)}_I$ at
each location, and the crosses indicating exact two-photon
resonance.}
\end{figure}

Now to consider the spatial dispersion of the signal light, suppose
that the signal frequency is varying linearly with $x$: $\Delta_s =
S_{\mathrm{disp}} x$.  If we chose $ S_{\mathrm{disp}} =S_B
s_\mathrm{res}$ then we would be in perfect two-photon resonance
everywhere.  However, this results in no differential phase shift
across the spectrum of the pulse and thus, no delay. To achieve a
delay, one must choose the these two slopes to be slightly
mismatched, as diagrammed in Fig.~\ref{fig:widebandSchematic}.  So
long as  the local detuning at the edges of the cell (or interaction
area) $ |\Delta_s'|=(S_{\mathrm{disp}} - S_B s_\mathrm{res})(0.5
w_\mathrm{int}) < \beta = W/\sqrt{D}$, the signal light is
everywhere locally in the EIT regime. Then we recover a differential
phase shift across the spectrum of the pulse, characterized by the
effective slope:

\begin{equation}
\label{eq:Seff} S_\mathrm{eff} = \frac{(S_{\mathrm{disp}} - S_B
s_\mathrm{res})}{S_{\mathrm{disp}}}S.
\end{equation}

\noindent If one chooses the maximum allowed mismatch
$\delta^\mathrm{(max)}$ then ratio by which are delay decreases
$S_{\mathrm{eff}}/S$ is the inverse of the ratio by which our
bandwidth increases $\beta_{\mathrm{eff}}/\beta$.

We note that the AC Stark shift in Scheme A must be accounted for to
properly choose the mismatch in frequencies.  However, the
differential AC Stark shift is linear with frequency and so can be
compensated for if needed.

\subsection{\label{subsec:resultsGradient} Results}
\begin{figure}
   \begin{center}
   \begin{tabular}{c}
   \includegraphics[height=2.5cm]{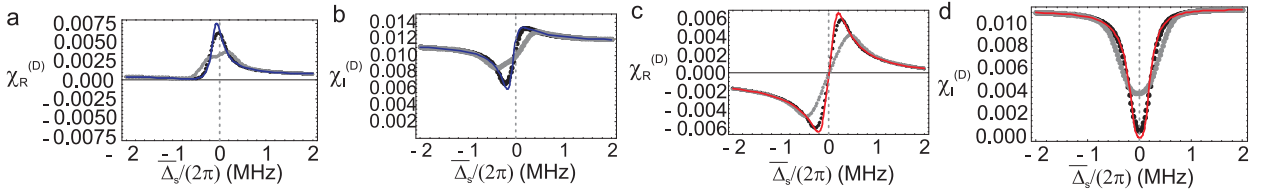}
     \end{tabular}
   \end{center}
\caption{\label{fig:chiDriftScheme}\textbf{Effect of magnetic field
gradient on the EIT resonance.}  \textbf{(a-b)} The real (a) and
imaginary (b) parts of $\chi^{(D)}$ for Scheme A with a pump power
5~mW and interaction area $w_\mathrm{int}=2$~cm,
$l_\mathrm{int}=0.5$~mm, and pressure $p=25$~Torr.  The solid (blue
online) curves show the resonance without any magnetic field
gradient, while the black and gray dots are calculated with
$S_B=2~$G/mm and $S_B=8~$G/mm, respectively, at the point where
$B_z=0$. \textbf{(c-d)} The same calculation for Scheme B (red
online).}
\end{figure}

In Fig.~\ref{fig:chiDriftScheme} we present examples of {\it local}
susceptibilities at $x=0 \, (B_z=0)$, calculated with the above
procedure in both Schemes A and B and with various magnetic field
gradients. The solid blue and red curves show the case with no
gradient while the black and gray dots show the results with the
gradients $S_B=2~$G/mm and $S_B=8~$G/mm, respectively.  As expected
the higher gradients wash out the EIT resonance, reducing both the
transparency and the slope.  This leads to a natural trade-off
between obtaining a higher bandwidth $\beta_{\mathrm{eff}}$ (with
larger gradients) and better transparency (with lower gradients). As
in the homogenous case, Scheme B offers better EIT for any given
$S_B$, though it is worth noting that the transparency in Scheme A,
because of the problem already present with level $\ket{4}$, is much
less sensitive to the introduction of gradients.

\begin{figure}
   \begin{center}
   \begin{tabular}{c}
   \includegraphics[height=2.5cm]{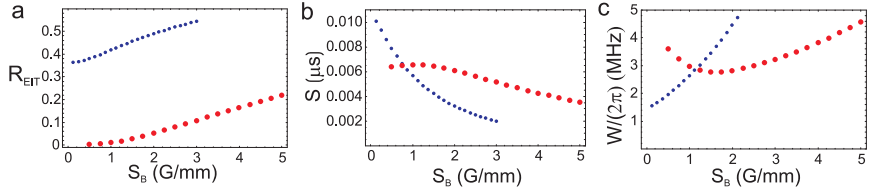}
     \end{tabular}
   \end{center}
\caption{\label{fig:pulseDrift}\textbf{Dependence of EIT resonance
on gradient.} \textbf{(a)} The relative EIT absorption
$R_\mathrm{EIT}=A/\chi^{(D)}(\Delta_0)|_{\Omega_p\rightarrow 0}$
versus magnetic field gradient in Scheme A (smaller, blue dots) and
Scheme B (larger, red dots). We use a pump power 5~mW and
interaction area $w_\mathrm{int}=2$~cm, $l_\mathrm{int}=0.5$~mm and
a pressure $p=10$~Torr for Scheme A and $p=25$~Torr Scheme B.
\textbf{(b)} The slope $S$ of the resonance for the same parameters.
\textbf{(c)} The width of the resonance $W$.}
\end{figure}

An analytic treatment calculating the correction
Eq.~(\ref{eq:rhoDiff}) using the non-Doppler broadened $\chi$
obtained from Eq.~(\ref{eq:rhoSS}), reveals that the expected
absorption at the resonance (in the absence of other decoherence
mechanisms from $\ket{4}$ and $\gamma_{\mathrm{diff}}$) is
$A_{\mathrm{grad}}=64(D_g/3)S_B^2 S_{\mathrm{res}}^2 \Gamma_r
\gamma_e^2/|\Omega_p|^6$.  We performed full numerical calculations
of $\chi^{(D)}$, using Eq.~(\ref{eq:rhoDiffInt}), choosing values of
$h_\mathrm{int}=0.5$~mm and $w_\mathrm{int}=2$~cm and a pump power
5~mW, which gives $\Omega_p=(2 \pi)12.0$~MHz for Scheme A and
$\Omega_p=(2 \pi)16.9$~MHz for Scheme B.  In
Fig.~\ref{fig:pulseDrift} we show the dependence of the absorption
$A$, slope $S$, and width $W$ versus $S_B$. The values chosen for
the pressure correspond to optimal choices we discuss later.  Our
analytic expression above provides a reasonable estimate but is not
quantitatively accurate.  In particular, note the predicted
quadratic dependence of the absorption $A_{\mathrm{grad}} \sim
S_B^2$ holds only for very small values of $S_B$ then it becomes
close to a linear dependence. As expected, the gradient contribution
is dominant in Scheme B even for very small values of $S_B$, while
in Scheme A the $\ket{4}$ contribution is dominant until about
$S_B=4$~G/mm. One also notes the gradient substantially impacts the
slopes $S$.

The performance as a delay device can be characterized again by the
maximum delay with $1/e$ attenuation.  For this analysis will choose
the maximum optical density $D_\mathrm{max}=1/2A$ and then demand
that the maximum detuning at the cell edges, $(S_{\mathrm{disp}} -
S_B s_\mathrm{res}) (0.5 w_\mathrm{int}) =
\delta^{\mathrm{(max)}}=W/\sqrt{2 D_\mathrm{max}}$ \footnote{The
maximum mismatch $\delta^{\mathrm{(max)}}$ is also restricted by the
fact that it must be in the region where there is a linear slope in
$\chi^{(D)}_\mathrm{int}$ to prevent distortion. The absorption
requirement $W/\sqrt{2 D_\mathrm{max}}$ is always more stringent
anyway, but for Scheme A, where $D_\mathrm{max}$ can be
substantially smaller, the distortion issue can become important and
we have accounted for this in our calculations by requiring that the
$\delta^{\mathrm{(max)}}$ is also in the region of the EIT
resonance.}. This results in an effective slope $S_\mathrm{eff}=S
\sqrt{2A}W/\beta_\mathrm{eff}$ and a delay-bandwidth product
$(S_\mathrm{eff} D_\mathrm{max}/2) \beta_\mathrm{eff}=W S/\sqrt{2
A}$, virtually identical to conventional slow light. Examination of
Fig.~\ref{fig:pulseDrift} shows that the increased absorption $A$
and decreased slope $S$ in fact decrease this product with
bandwidth.

\begin{figure}
   \begin{center}
   \begin{tabular}{c}
   \includegraphics[height=2.5cm]{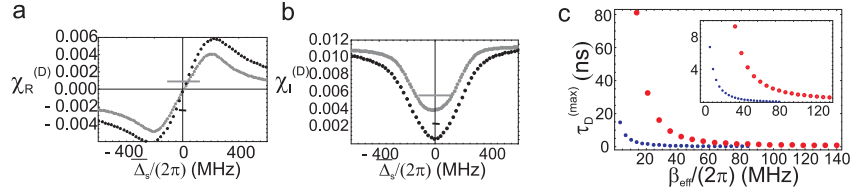}
     \end{tabular}
   \end{center}
\caption{\label{fig:wideband}\textbf{Dependence of EIT resonance on
gradient and bandwidth.} \textbf{(a,b)} The resulting effective
susceptibility for Scheme B, with with parameters as in
Fig.~\ref{fig:chiDriftScheme} ($S=2$~G/mm (black) and  $S=8$~G/mm
(gray)), when one chooses the mismatch such that
$(S_\mathrm{disp}-S_B S_{\mathrm{res}})w_\mathrm{int}/2=
\delta^{\mathrm{(max)}}$.  The horizontal bars (black and gray)
indicate the bandwidth $\beta_{\mathrm{eff}}$.  \textbf{(c)} The
maximum delay $\tau_D^{\mathrm{(max)}}=S_\mathrm{eff}
D_\mathrm{max}/2$ for various gradients $S_B$ versus the resulting
bandwidth $\beta_{\mathrm{eff}} = s_\mathrm{res} S_B w_\mathrm{int}$
in Scheme A (smaller, blue) and B (larger, red).  We have kept the
pump power 5~mW and the interaction area $w_\mathrm{int}=2$~cm, $
h_\mathrm{int}=0.5$~mm constant, and chosen $p=10$~Torr for Scheme A
and $p=25$~Torr for Scheme B.  The inset shows a zoom in on the
smaller delay points.}
\end{figure}

In Figs.~\ref{fig:wideband}(a-b), we show the effective dispersion
curves for Scheme B the cases in Fig.\ref{fig:chiDriftScheme}(c-d),
with the mismatch chosen to be the maximum allowed difference as
just described. These curves are just stretched in $\Delta_s$, with
the stretch factor being the inverse of the ratio on the right-hand
side of Eq.~(\ref{eq:Seff}).  The black and gray horizontal bars
indicate the bandwidths $\beta_\mathrm{eff}$ (the frequencies
dispersed within the $w_\mathrm{int}=2$~cm area) in each case.  In
Fig.~\ref{fig:pulseDrift}(c) we plot the maximum delay versus the
bandwidth in both schemes. It appears delays $\sim 5$~ns are
possible over a bandwidth $\sim 50$~MHz in Scheme B, whereas similar
delays are only possible over  $\sim 10$~MHz in Scheme A.

\begin{figure}
   \begin{center}
   \begin{tabular}{c}
   \includegraphics[height=2.5cm]{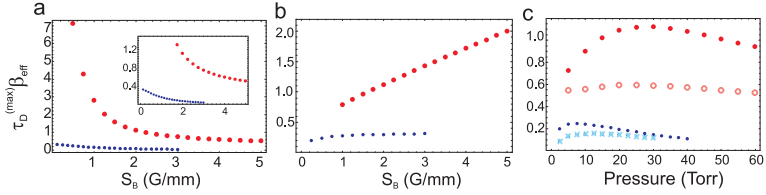}
     \end{tabular}
   \end{center}
\caption{\label{fig:parameterDrift}\textbf{Dependence of delay
capabilities on field gradient buffer gas pressure.} \textbf{(a)}
The maximum delay bandwidth product $\tau_D^{\mathrm{(max)}}
\beta_{\mathrm{eff}}$ versus the slope $S_B$ for the same parameters
as in Fig.~\ref{fig:wideband}(c).  The inset shows a zoom in on the
lower delay-bandwidth product points. Note that $S_B=$1~G/mm
corresponds to $\beta_\mathrm{eff}=(2 \pi)23$~MHz.  \textbf{(b)}
Same plot, but adjusting the width $w_\mathrm{int}$ so the bandwidth
is constant, keeping the total pump power at 5~mW. The Scheme B
series is chosen so that the total magnetic field difference across
$w_\mathrm{int}$ is 40~G ($\beta_\mathrm{eff}=(2 \pi)56$~MHz) and
for the Scheme A series is 10~G ($\beta_\mathrm{eff}=(2
\pi)14$~MHz). \textbf{(c)} Delay/bandwidth product versus pressure
keeping $w_\mathrm{int}=2$~cm.  The solid, large (red online) dots
are Scheme B for $S_B=2$~G/mm and the non-solid (light red) circles
are for $S_B=4$~G/mm.  For Scheme A for $S_B=0.5$~G/mm (small, dark
blue dots) and $S_B=1$~G/mm (x's, light blue). }
\end{figure}

To get a better sense of the sacrifices one makes to get a wider
bandwidth (due to the diffusion problem) we plot in
Fig.~\ref{fig:parameterDrift}(a)  the maximum delay-bandwidth
product versus magnetic field gradient (with the same parameters as
Fig.~\ref{fig:wideband}(c)).  We see that for Scheme B we maintain
full pulse separation capabilities up to bandwidths
$\beta_\mathrm{eff} \sim 50$~MHz.

Because of the extremely strong dependence of $A_\mathrm{grad}$ with
power we found numerically that one always benefits from using
smaller $h_\mathrm{int}$, so this parameter should be chosen to be
the smallest reasonable value over which the beam can be easily
focused.  Though too small a value $h_\mathrm{int}$ would lead to a
higher loss from $\gamma_\mathrm{diff}$ this problem is almost
always dominated by the diffusion in the magnetic field gradient and
so is not a big consideration. Similarly, in choosing
$w_\mathrm{int}$ we found that the gain in transparency from higher
intensities tended to outweigh the gain one got from lower slopes.
Figure~\ref{fig:parameterDrift}(b) shows the maximum delay-bandwidth
product versus $S_B$, but keeping the pump power constant and
adjusting the interaction with $w_\mathrm{int}$ such that the
bandwidth $\beta_\mathrm{eff}=(2 \pi) 56$~MHz was also kept
constant. Indeed one sees that one gains by using steeper gradients
over smaller areas.  Ultimately, the slope $S_B$ that can be used in
practice will be determined by the manner in which the magnetic
field gradient and signal dispersion can be generated.

In the calculations in Figs.~\ref{fig:wideband} and
\ref{fig:parameterDrift}(a-b) we chose pressures $p=10$~Torr and
25~Torr for Schemes A and B, respectively. Numerically these were
found to be about optimal. In Fig.~\ref{fig:parameterDrift} we plot
the dependence of the of the maximum delay-bandwidth on the pressure
for several cases.  In Scheme A, the optimal pressures are larger
than in the homogenous case due to increased importance of reducing
diffusion in the magnetic field gradient. Interestingly, even in
Scheme B, higher pressures eventually reduce the performance. This
can be understood from the factor $\gamma_e^2$ in the analytic
estimate for $A_\mathrm{grad}$ above. The physical origin of this
factor is the fact that the EIT width $W=\Omega_p^2/\sqrt{8\gamma_e
\Gamma_r}$ decreases with $\gamma_e$ and therefore makes the
resonance more sensitive to the averaging over nearby magnetic
fields Eq.~(\ref{eq:rhoDiffInt}). We plot the dependence on pressure
for two different gradients in each scheme. While the optimal
pressures are slightly different, the dependence on pressure is
rather weak and so a sensitive parameter search versus $p$ should
not be required.

\section{\label{sec:conclusion} Summary}

We have performed a comprehensive and systematic analysis of EIT
resonances, and the resulting pulse propagation characteristics, in
${}^{87}$Rb vapors, including effects of couplings to additional
levels in the hyperfine structure and a buffer gas.  We then
calculated the delays, transmissions, and bandwidths for propagation
of light tuned to these resonances.   We analyzed two particular
$\Lambda$ level schemes (diagrammed in Fig.~\ref{fig:diagram}) and
found that Scheme B was far superior, in terms of achievable delays
and delay-bandwidth products, due to the lack of coupling to
additional levels. Despite its poorer performance, Scheme A still
provides reasonable performance and may be desirable since it is
much easier to initialize, simply with optical pumping. Importantly,
we found the EIT resonance could be shifted over a wide range of
frequencies by applying a homogeneous magnetic field, and that the
resonance characteristics were quite insensitive to this field over
range of about 500~MHz. This analysis serves as a useful model to
study EIT in conventional slow light, and also as a basis for study
of our channelization architecture.

We then presented a model to analyze the effect of an inhomogenous
magnetic field, which causes a strong variation of the EIT resonance
frequency in the transverse direction.  This was then applied to
analyze the performance of our proposed channelization architecture
for wide-band slow light, where a signal pulse is spatially
dispersed according to frequency and an inhomogeneous magnetic field
is applied in a such a way that the EIT resonance frequency matches
this dispersion. We found that by choosing the magnetic field
gradient so the change in the two-photon resonance is slightly
mismatched from the transverse dispersion of the signal, one could
achieve EIT and slow light conditions over a much larger bandwidth
than with conventional slow light.  This is essential for
applications in many signal processing applications. We found that
the diffusion of atoms in the field tended to reduce the
delay-bandwidth products with bandwidth.  In Scheme B, this
architecture should allow a delay-bandwidth greater than unity up to
bandwidths of about $\sim 50$~MHz, where delays are $\sim 5$~ns (see
Fig.~\ref{fig:pulseDrift}). Furthermore, we note either the pump
field power or the magnetic field gradient can be used to control
$S_{\mathrm{eff}}$ and thus the delay, making it a controllable time
delay system.

The buffer gas is important in reducing the diffusion of atoms from
into regions of widely varying magnetic field and so higher buffer
gas pressures are generally desirable for higher magnetic field
gradients. However, we also found that higher pressures narrow the
EIT feature and can therefore increase the sensitivity of the
dispersive slope and absorption profile to magnetic field gradients.
Balancing these two considerations leads to an optimal pressure,
which we found this optimum to be near $p\sim$10~Torr for Scheme A
and $p \sim 25$~Torr for Scheme B, for reasonable parameters.   This
optimal pressure was not very sensitive to the exact value of the
gradient and other parameters.  We also found that one generally
benefited from tight focusing and high magnetic gradients.

In future work, it will be useful to consider the effects of atomic
diffusion at a more microscopic level.  In particular, it has been
found that the model used here for diffusion out of the interaction
region may overestimate the loss in real systems due to the fact
that atoms can diffuse back into the interaction region
\cite{Irina}.  Additionally, dynamical jumps of velocity of
individual rubidium atoms upon collisions with the buffer gas has
also been found to be an important consideration \cite{bufferRot}.
Finally, for implementation of this system, work is also needed to
develop optimal methods for transversely dispersing the signal field
and producing large linear magnetic field gradients.

The role played by the differential phase shift in this system is
interesting in its own right and merits further investigation. It is
not entirely clear that the pulse will not be significantly more
slowed than our analysis here shows, due to subtleties with the
transverse dispersion of the signal.  Perhaps the signal dispersion
or magnetic field could be engineered in such a way that the group
velocity is governed by the {\it local} (and much larger) $S$,
rather than $S_{\mathrm{eff}}$, allowing much larger delays.
Furthermore, it may be possible to combine this method with aspects
of previous light storage experiments
\cite{stoppedCold,stoppedThermal} to significantly increase the
delay times.

%%%%%%%%%%%%%%%%%%%%%%%%%%%%%%%%%%%%%%%%%%%%%%%%%%%%%%%%%%%%%
\acknowledgments

The authors wish to thank Irina Novikova, Mikhail Lukin, and Fredrik
Fatemi for helpful discussions.  This work was supported by the
Office of Naval Research and the Defense Advanced Research Projects
Agency (DARPA) Slow Light Program.

%%%%%%%%%%%%%%%%%%%%%%%%%%%%%%%%%%%%%%%%%%%%%%%%%%%%%%%%%%%%%
%%%%% References %%%%%

\end{document}